\documentclass[11pt,epsf]{article}
\usepackage{graphicx}
\usepackage{color}
\usepackage{amsmath}
\usepackage{amsfonts}
\usepackage{mathrsfs}
\usepackage{mathtools}
\usepackage{amssymb}
\usepackage{float}
\usepackage{url}
\usepackage{verbatim}
\usepackage{dsfont}
\usepackage{enumitem}
\usepackage{layout}
\usepackage{comment}
\setlist[enumerate]{itemsep=0mm}
\setlist[enumerate]{align=left}
\setlist[itemize]{align=left}

\newcommand {\dfn} {\stackrel{\Delta} {=}}

\newcommand{\eqa}{\stackrel{\mbox{(a)}}{=}}

\newcommand{\gea}{\stackrel{\mbox{(a)}}{\ge}}
\newcommand{\geb}{\stackrel{\mbox{(b)}}{\ge}}

\newcommand {\reals} {{\rm I\!R}}

\newcommand {\bsigma} {\mbox{\boldmath $\sigma$}}

\newcommand {\bu} {\mbox{\boldmath $u$}}
\newcommand {\bv} {\mbox{\boldmath $v$}}

\newcommand {\by} {\mbox{\boldmath $y$}}
\newcommand {\bz} {\mbox{\boldmath $z$}}

\newcommand {\bE} {\mbox{\boldmath $E$}}

\newcommand{\calA}{{\cal A}}
\newcommand{\calB}{{\cal B}}

\newcommand{\calD}{{\cal D}}

\newcommand{\calI}{{\cal I}}

\newcommand{\calS}{{\cal S}}

\newcommand{\calU}{{\cal U}}
\newcommand{\calV}{{\cal V}}

\newcommand{\calY}{{\cal Y}}
\newcommand{\calZ}{{\cal Z}}
\allowdisplaybreaks

\begin{document}
\thispagestyle{empty}
\title{Refinements and Generalizations of the Shannon Lower Bound via
Extensions of the Kraft Inequality}

\author{Neri Merhav
%\thanks{
%Currently on sabbatical leave at HP Laboratories,
%1501 Page Mill Road, MS 3U-4, Palo Alto CA 94304, USA.}
}
\date{}
\maketitle

\begin{center}
The Andrew \& Erna Viterbi Faculty of Electrical Engineering\\
Technion - Israel Institute of Technology \\
Technion City, Haifa 32000, ISRAEL \\
E--mail: {\tt merhav@ee.technion.ac.il}\\
\end{center}
\vspace{1.5\baselineskip}
\setlength{\baselineskip}{1.5\baselineskip}

\begin{abstract}
We derive a few extended versions of the Kraft inequality for
lossy compression, which pave the way to the derivation of several
refinements and extensions of the well known Shannon lower bound in a variety
of instances of rate-distortion coding. These refinements and extensions
include sharper bounds for one-to-one codes and $D$-semifaithful codes, a
Shannon lower bound
for distortion measures based
on sliding-window functions, and an individual-sequence counterpart of the
Shannon lower bound.
\end{abstract}

\setcounter{section}{0}

\section{Introduction}
\label{intro}

The Shannon lower bound (SLB) is one of the most important analytic tools in
rate-distortion theory because it provides a simple, explicit, 
and often very tight lower bound to the rate-distortion function for 
a wide class of sources and distortion measures, see, e.g., Subsection 3.4.1
of \cite{Berger71}, Sections 4.3 and 4.6 of \cite{Gray90}, and Problem 10.6 of
\cite{CT06}. Its significance lies in giving a universal benchmark that connects 
rate-distortion tradeoffs to the entropy or the differential entropy of the
source (depending on whether the source has a discrete or continuous alphabet), 
thereby offering an intuitively transparent approximation in regimes 
where exact evaluation of the rate-distortion function is intractable. 
The SLB is particularly powerful at low distortion, where it frequently 
coincides with the exact rate-distortion function for smooth sources, and it serves as a 
foundation for many refinements and asymptotic approximations 
(e.g., high-resolution analysis, corrections for lattice quantizers, and
recent non-asymptotic bounds). In fact, the literature contains reported
results on the asymptotic tightness of the SLB in the limit of low distortion
under fairly mild regularity conditions \cite{LZ94}, \cite{Koch16}.
Because computing the rate-distortion function
exactly is generally difficult, the SLB plays a central role in analysis, design, 
and performance assessment of lossy compression schemes. More recent studies
include the finite block-length regime \cite{Kostina15}, \cite{Kostina16} and
further developments concerning the quadratic distortion function
\cite{GS24}.

In \cite{Campbell73} Campbell derived an extension of Kraft's inequality, that
leads to the SLB in both the discrete- and the continuous alphabet
cases. However, his results were claimed to apply to rate-distortion codes whose distortion
level is defined by the distance between the two most distant source-space
vectors that are mapped to the same codeword, namely, the diameter (as opposed to the radius) of
the distortion ball centered at the reproduction vector. In
\cite{me95}, a similar Kraft inequality was derived for $D$-semifaithful
codes, namely, codes that incur per-letter distortion that never exceeds $D$
in the usual sense. An additional benefit of the derivation in \cite{me95} was
that it could deliver also an $O\left(\frac{\log n}{n}\right)$ redundancy term
on top of the SLB at least for certain distortion measures for
which there is an explicit expression of the cardinality (or the volume, in
the continuous case) of a ball of normalized radius $D$ 
in the source vector space.

In this work, we propose a few other extended versions of Kraft's inequality that
together pave the way to several further refinements and generalizations of the
SLB. In these extensions of Kraft's inequality, the idea is to upper bound
the summation (or the integral, in the continuous case) of an exponentiated
negative linear combination of the code-length and the distortion incurred by
each and every vector in the source space. By contrasting an upper bound with
a lower bound to this quantity, we obtain
several refinements and extensions to the SLB, which apply to the following
scenarios.\\

\noindent
1. {\em One-to-one codes.} Instead of assuming that the reproduction vectors
are represented by uniquely decodable (UD) codes, we relax this restriction
and allow any one-to-one code in the level of $n$-vectors. The lower bound
then becomes the SLB minus an $O\left(\frac{\log n}{n}\right)$ term, similarly
as in the lossless case derived by Rissanen \cite{Rissanen82}.\\

\noindent
2. {\em $D$-semifaithful codes.} Similarly as in \cite{me95}, we consider
$D$-semifaithful codes, but here we allow several simultaneous distortion
criteria. Using saddle-point integration, we show that the resulting lower
bound is given by the SLB plus $\frac{k\log n}{2n}+o\left(\frac{1}{n}\right)$,
where $k$ is the ``effective number'' of distortion constraints. The concrete
meaning of this term will be provided in the sequel.\\

\noindent
3. {\em Sliding-window distortion functions.} In some applications, one might
wish to shape the spectrum or the memory properties of the reconstruction error signal. This can be
done by imposing additional distortion constraints defined by additive functions that operate on two or more consecutive
samples of the error signal in sliding-window fashion. Our framework is
capable of incorporating such distortion functions and allowing a derivation
of generalized SLB for this case.\\

\noindent
4. {\em Individual sequences and finite-state encoders.}
By developing a generalized Kraft inequality, similar (but not identical) to
the one by Ziv and
Lempel \cite{ZL78} for finite-state encoders, we derive also an individual-sequence counterpart of the
SLB, where the source entropy term is replaced by the
Lempel-Ziv complexity.\\

As is well known, the classical SLB can be obtained
significantly more simply and easily than going via the Kraft inequality. In particular, 
it is obtained by a straightforward direct manipulation of the mutual
information. But it should be emphasized that the point of this article is not in a quest for a simpler proof of
the SLB. The point is that the path that goes via the Kraft inequality leads
to the above mentioned extensions and the refinements.

The outline of the remaining part of this article is as follows.
In Section \ref{ncb}, we establish some notation conventions (Subsection
\ref{nc}) and provide elementary  background on the SLB (Subsection \ref{bg}).
In Section \ref{extendedkraft}, we present and prove our extended Kraft inequality in
several variations. In Section \ref{lowerbounds}, we derive corresponding
lower bounds, first, for UD lossless compression of the reproduction data,
then for one-to-one compression thereof (Subsection \ref{121}), and finally,
for $D$-semifaithful codes (Subsection \ref{Dsf}). In Section
\ref{slidingwindow}, we
address the case of sliding-window distortion functions. Finally, in Section
\ref{indivseq}, we first provide some background on finite-state compression
of individual sequences and the LZ algorithm (Subsection \ref{indivbg}) and
then derive an individual-sequence counterpart of the SLB (Subsection
\ref{indivslb}).

\section{Notation Conventions and Background}
\label{ncb}

\subsection{Notation Conventions}
\label{nc}

Throughout this paper, scalar random
variables (RV's) will be denoted by capital
letters, their sample values will be denoted by
the respective lower case letters, and their alphabets will be denoted
by the respective calligraphic letters.
A similar convention will apply to
random vectors and their sample values,
which will be denoted with same symbols superscripted by the dimension.
Thus, for example, $U^n$ ($n$ -- positive integer)
will denote a random $n$-vector $(U_1,...,U_n)$,
and $u^n=(u_1,...,u_n)$ is a specific vector value in $\calU^n$,
the $n$--th Cartesian power of $\calU$, which is the alphabet of each
component of $u^n$. In some of our derivations below,
there will be a need to refer to multiple copies of the vector $u^n$. In such
cases, in order to avoid cumbersome subscripts and superscripts for indexing,
we will use the alternative notation $\bu$ for $u^n$, and then the various
copies will be denoted by $\bu_1$, $\bu_2$, etc. Returning to the first
notation method,
$u_i^j$ and $U_i^j$, where $i$
and $j$ are integers and $i\le j$, will designate segments $(u_i,\ldots,u_j)$
and $(U_i,\ldots,U_j)$, respectively,
where for $i=1$, the subscript will be omitted (as above).
For $i > j$, $u_i^j$ (or $U_i^j$) will be understood as the null string.
Logarithms and exponents, throughout this paper, will be understood to be taken to the base 2
unless specified otherwise. The indicator of an event $\calA$ will be denoted
by $\calI\{\calA\}$, i.e., $\calI\{\calA\}=1$ if $\calA$ occurs and
$\calI\{\calA\}=0$ if not.

Sources and probability distributions associated with them will be denoted generically by the letter $P$
subscripted by the name of the RV and its conditioning,
if applicable, exactly like in
ordinary textbook notation standards, e.g., $P_{U^n}(u^n)$ is the probability function of
$U^n$ at the point $U^n=u^n$, $P_{X|U^n}(x|u^m)$
is the conditional probability of $X=x$ given $U^n=u^n$, and so on.
Whenever clear from the context, these subscripts will be omitted.
Information theoretic quantities, like entropies and mutual
informations, will be denoted following the usual conventions
of the information theory literature, e.g., $H(U^n)$, $I(S;U^n|V^n)$,
and so on. The differential entropy of a continuous valued RV, $U^n$, will be
denoted by $h(U^n)$. The expectation operator will be denoted by
$\bE\{\cdot\}$ and the probability of an event $\calA$ will be denoted by
$\mbox{Pr}\{\calA\}$.

It should be noted that our derivations will apply to both discrete-alphabet
sources and to continuous alphabet sources. To avoid repetitions, we
henceforth carry on under the assumption of a continuous alphabet source with
the understanding that in the discrete-alphabet cases, integrations over
$\calU^n$ should simply be replaced by summations.

Let $U^n=(U_1,U_2,\ldots,U_n)$ denote a source vector, drawn from a stochastic
process, $P$, whose alphabet is $\calU$. The source vector is compressed by a lossy fixed-to-variable (F-V) source
code, defined by an encoder $\phi_n:\calU^n\to\calB_n\subset\{0,1\}^\star$ and a decoder
$\psi_n:\calB_n\to\calV_n\subseteq\calU^n$, where $\calB_n$ denotes a certain subset of the
set of all binary variable-length strings, $\{0,1\}^*$. Without loss of
generality (and optimality), it is assumed that the
encoder $\phi_n$, can be viewed as a cascade of a
reproduction encoder (vector quantizer) $\calU^n\to\calV_n$ and a uniquely
decodable (UD) lossless code $\calV_n\to\calB_n$. 
In a certain part of our
results this unique decodability assumption will be partially relaxed to become
the less demanding assumption of a one-to-one mapping. Let $L[\phi_n(u^n)]$
denote the length (in bits) of the compressed codeword, $\phi_n(u^n)$.

Assuming that $\calU$
is a group with certain addition and subtraction operations (e.g.,
modulo-$K$ addition/subtraction for $\calU=\{0,1,\ldots,K-1\}$ for a finite
positive integer $K$, or ordinary addition/subtraction for $\calU=\reals$)
we will focus on additive difference distortion measures, where the distortion
between the source vector $u^n\in\calU^n$ and 
the reproduction vector $v^n=\psi_n(\phi_n(u^n))\in\calV_n$ will be given by 
\begin{equation}
d(u^n,v^n)=\sum_{i=1}^nd(u_i,v_i)=\sum_{i=1}^n\rho(u_i-v_i),
\end{equation}
where $\rho(z)$, $z\in\calU$ is a certain non-negative function, 
which vanishes if and only if $z=0$, for example,
$\rho(z)=|z|$, or $\rho(z)=z^2$, etc. For the sake of convenience, we will
sometimes denote $d(u^n,v^n)$ by $\rho(u^n-v^n)$, which for a given
encoder-decoder pair, is also $\rho(u^n-\psi_n(\phi_n(u^n)))$. 
Given an encoder-decoder pair, $(\phi_n,\psi_n)$, the expected distortion,
$\sum_{i=1}^n\bE\{\rho(U_i-V_i)\}$ will be constrained to be less than or
equal to $nD$, where $D>0$ designates the per-letter distortion level allowed.
In certain parts of our derivations, the more restrictive pointwise distortion
constraint will be imposed, i.e., $\rho(u^n-\psi_n(\phi_n(u^n)))\le nD$ for
all $u^n\in\calU^n$. In other parts,
more than one difference distortion measure will play a
role, and accordingly, more than one distortion constraint will be imposed.
Given $k$ difference distortion functions, $\rho_j(\cdot)$, $j=1,2\ldots,k$,
we then require $\bE\{\rho_j(U^n-\psi_n(\phi_n(U^n)))\}\le nD_j$ (or
$\max_{u^n\in\calU^n}\rho_j(u^n-\psi_n(\phi_n(u^n)))\}\le nD_j$)
for all $j=1,2,\ldots,k$.
In another part of our results, we also allow distortion functions to be
additive sliding-window functions operating on $m$ consecutive symbols of the
difference $z^n=u^n-v^n$. I.e.,
\begin{equation}
\rho(z^n)=\sum_{i=m}^n\rho(z_{i-m+1}^i).
\end{equation}
This corresponds to situations where we wish to shape not only the `intensity'
of the error signal, $z^n$, but also its memory properties, for example, the
correlations between consecutive symbols of $z^n$. We will elaborate more on
this in Section \ref{slidingwindow}.

Similarly as
with the $u^n$, for which we adopt the alternative notation $\bu$, the same
will apply to $v^n$, which will denoted also by $\bv$, along with its multiple
copies $\bv_1$, $\bv_2$, and so on.

\subsection{Background}
\label{bg}

For a continuous-alphabet memoryless source $P$, the SLB is given by
\begin{equation}
R(D)\ge R_{\mbox{\tiny SLB}}(D)\dfn h(U)-\Phi(D),
\end{equation}
where $h(U)$ is the differential entropy of a single symbol $U$ and
\begin{equation}
\label{equivalence}
\Phi(D)\dfn\sup_{\{Z:~\bE\{\rho(Z)\}\le D\}}h(Z)=\inf_{\beta\ge 0}\left\{\beta
D+\log\left[\int_{\calU}2^{-\beta\rho(z)}\mbox{d}z\right]\right\},
\end{equation}
assuming that $\int_{\calU}2^{-\beta\rho(z)}\mbox{d}z<\infty$ for some
$\beta>0$. The
equivalence between the two expressions of $\Phi(D)$ can be
easily shown using standard techniques. For the sake of completeness, we prove this equivalence in the
appendix (see also Section 4.3.1 in \cite{Berger71} and Sections 4.3 and 4.6
in \cite{Gray90}). The advantage of the second formula of $\Phi(D)$ is that it involves
optimization over one parameter only, as opposed to variational calculus in
the first formula. Clearly, if the source is discrete rather than continuous,
the differential entropy, $h(U)$, should be replaced by ordinary entropy,
$H(U)$, and the integration over $\calU$ should be replaced by summation, as
indicated above. 

For a source with memory,
the SLB is given by
\begin{equation}
R_n(D)\dfn\min_{\{P_{V^n|U^n}:~\bE\{\rho(U^n-V^n)\le
nD\}}\frac{I(U^n;V^n)}{n}\ge\frac{h(U^n)}{n}-\Phi(D),
\end{equation}
which in the limit of large $n$. becomes
\begin{equation}
R(D)\ge \lim_{n\to\infty}\frac{h(U^n)}{n}-\Phi(D)\dfn \bar{h}(U^\infty)-\Phi(D),
\end{equation}
where $\bar{h}(U^\infty)$ is the differential entropy rate of the source.
In all cases, the function $\Phi(D)$ remains as in (\ref{equivalence}).

\section{Extended Kraft Inequalities}
\label{extendedkraft}

For a given encoder-decoder pair, $(\phi_n,\psi_n)$, and parameters
$\alpha>1$ and $\beta\ge 0$, define the {\em extended Kraft integral} (or, the
extended Kraft
sum, in the discrete-alphabet case) as:
\begin{equation}
Z^n(\alpha,\beta)\dfn\int_{\calU^n}\exp_2\{-\alpha
L[\phi_n(u^n)]-\beta\rho(u^n-\psi_n(\phi_n(u^n)))\}\mbox{d}u^n.
\end{equation}

\noindent
{\em Lemma 1.}
Let $(\phi_n,\psi_n)$ induce a UD lossless compression of
$v^n=\psi_n(\phi_n(u^n))$.
Then, for every $\alpha>1$ and $\beta\ge 0$,
\begin{equation}
Z^n(\alpha,\beta)\le 
\left[\int_{\calU} 2^{-\beta\rho(z)}\mbox{d}z\right]^n.
\end{equation}

\noindent
{\em Proof of Lemma 1.}
Let us examine the expression of $[Z_n(\alpha,\beta)]^k$ for an arbitrary positive
integer $k$. 
\begin{eqnarray}
[Z^n(\alpha,\beta)]^k&=&\left[\int_{\calU^n}\exp_2\{-\alpha
L[\phi_n(\bu)]-\beta\rho(\bu-\psi_n(\phi_n(\bu)))\}\mbox{d}\bu\right]^k\nonumber\\
&=&\int_{\calU^n}\mbox{d}\bu_1\cdot\cdot\cdot\int_{\calU^n}\mbox{d}\bu_k\cdot
\exp_2\left\{-\alpha\sum_{i=1}^kL[\phi_n(\bu_i)]-\beta\sum_{i=1}^k\rho(\bu_i-\psi_n(\phi_n(\bu_i)))\right\}\nonumber\\
&\le&\sum_{\ell=1}^\infty\sum_{\{\{\bv_i\}_{i=1}^k:~\sum_{i=1}^kL(\bv_i)]=\ell\}}
\int_{\{\bu_1:~\psi_n(\phi_n(\bu_1))=\bv_1\}}\mbox{d}\bu_1\cdot\cdot\cdot
\int_{\{\bu_k:~\psi_n(\phi_n(\bu_k))=\bv_k\}}\mbox{d}\bu_k\times\nonumber\\
& &\exp_2\bigg\{-\alpha\ell-\beta\sum_{i=1}^k\rho(\bu_i-\bv_i)\bigg\}\nonumber\\
&\le&\sum_{\ell=1}^\infty
2^{-\alpha\ell}\sum_{\{\{\bv_i\}_{i=1}^k:~\sum_{i=1}^kL(\bv_i)=\ell\}}
\int_{\calU^n}\mbox{d}\bz_1\cdots
\int_{\calU^n}\mbox{d}\bz_k
\exp_2\left\{-\beta\sum_{i=1}^k\rho(\bz_i)\right\}\nonumber\\
&\le&\sum_{\ell=1}^\infty
2^{-\alpha\ell}\cdot 2^{\ell}\left[\int_{\calU^n}\mbox{d}z^n
\exp_2\{-\beta\rho(z^n)\}\right]^k\nonumber\\
&=&\left[\int_{\calU^n}\cdot\mbox{d}z^n
\exp_2\{-\beta\rho(z^n)\}\right]^k\cdot\sum_{\ell=1}^\infty
2^{-(\alpha-1)\ell}\nonumber\\
&=&\frac{\left[\int_{\calU}\mbox{d}z
\exp_2\{-\beta\rho(z)\}\right]^{nk}}{2^{\alpha-1}-1},
\end{eqnarray}
and so,
\begin{equation}
Z^n(\alpha,\beta)\le\frac{\left[\int_{\calU}\mbox{d}z
\exp_2\{-\beta\rho(z)\}\right]^n}{(2^{\alpha-1}-1)^{1/k}},
\end{equation}
which upon taking the limit $k\to\infty$, becomes
\begin{equation}
Z^n(\alpha,\beta)\le\left[\int_{\calU}
2^{-\beta\rho(z)}\mbox{d}z\right]^n,
\end{equation}
completing the proof of Lemma 1.\\

For one-to-one codes, instead of taking the limit of $k\to\infty$ in the proof
of Lemma 1, we simply set $k=1$, since the one-to-one property is merely
imposed in the level of a single $n$-block rather than in the level of concatenations
of blocks, as in UD codes. This yields the following variation of Lemma 1.\\

\noindent
{\em Lemma 2.}
Let $(\phi_n,\psi_n)$ induce a one-to-one mapping between
$\psi_n(u^n)$ and $v^n=\psi_n(\phi_n(u^n))$.
Then, for every $\alpha>1$ and $\beta\ge 0$,
\begin{equation}
Z_{\mbox{\tiny 1-1}}^n(\alpha,\beta)\le 
\frac{\left[\int_{\calU}2^{-\beta\rho(z)}\mbox{d}z\right]^n}{2^{\alpha-1}-1},
\end{equation}
where $Z_{\mbox{\tiny 1-1}}^n(\alpha,\beta)$ is defined exactly as
$Z^n(\alpha,\beta)$, except that it may apply to the larger class of
one-to-one codes, rather than UD codes.\\

When $\alpha\in(1,2)$, the denominator, $2^{\alpha-1}-1$, is smaller than
unity, and then the upper bound to $Z_{\mbox{\tiny 1-1}}^n(\alpha,\beta)$ in Lemma 2 is larger than that of
Lemma 1. Indeed, the interesting region for selecting the values of
$\alpha$ is in the vicinity of unity, where $2^{\alpha-1}-1<1$.

Returning to the class of UD lossless encodings of $v^n=\psi_n(\phi_n(u^n))$,
two additional variations of the above extended Kraft inequality can be
considered. The first pertains to $D$-semifaithful codes, namely, codes for
which the distortion is restricted to never exceed $nD$, pointwise, and not
merely in expectation, that is,
$\max_{u^n\in\calU^n}\rho(u^n-\psi_n(\phi_n(u^n)))\le nD$, where $D$ is the
allowed per-letter distortion. The second variation is
associated with fixed-rate codes, i.e., codes for which $L[\phi_n(u^n)]=nR$
for all $u^n\in\calU^n$, where $R>0$ is the allowed coding rate.

For $D$-semifaithful codes, let us re-define the integrand of the extended Kraft
integral by replacing
the term $\beta\rho(u^n-\psi_n(\phi_n(u^n)))$ at the exponent
with the function $W(\rho(u^n-\psi_n(\phi_n(u^n)))-nD)$, where $W[\cdot]$ is the
infinite well function (IWF),
\begin{equation}
\label{iwf}
W(t)\dfn\left\{\begin{array}{ll}
0 & t\le 0\\
\infty & t>0\end{array}\right.
\end{equation}
This causes the integrand of the extended Kraft integral to vanish wherever the
distortion constraint is violated, and then the extended Kraft integral
becomes
\begin{equation}
Z_{\mbox{\tiny D-sf}}^n(\alpha)\dfn\int_{\calS_n(D)}\exp_2\{-\alpha
L[\phi_n(u^n)]\}\mbox{d}u^n,
\end{equation}
where $\calS_n(D)=\{u^n:~\rho(u^n-\psi_n(\phi_n(u^n)))\le nD\}$. In this case, 
a simple modification of the proof of Lemma 1 yields the following version
of the extended Kraft inequality (see also
\cite{me95}):\\

\noindent
{\em Lemma 3.}
Let $(\phi_n,\psi_n)$ be a $D$-semifaithful code that comprises UD lossless encoding of
$v^n=\psi_n(\phi_n(u^n))$.
Then, for every $\alpha>1$,
\begin{equation}
Z_{\mbox{\tiny D-sf}}^n(\alpha)\le 
\mbox{Vol}\{z^n:~\rho(z^n)\le nD\}.
\end{equation}
\\

Here too, if the UD property is replaced by one-to-one property, then the right-hand side
should be divided by $2^{\alpha-1}-1$.

Finally, for fixed-rate codes, let us replace the term $L[\phi_n(u^n)]$ of the
Kraft integrand by $nR$ and return the second term therein to be
$\beta\rho(u^n-\psi_n(\phi_n(u^n)))$.
By similar manipulations of the proof of Lemma 1, we find that for fixed rate codes, the Kraft integral becomes
\begin{equation}
Z_{\mbox{\tiny fr}}^n(\alpha,\beta)\dfn \int_{\calU^n} \exp_2\{-\alpha
nR-\beta\rho(u^n-\psi_n(\phi_n(u^n)))\}\mbox{d}u^n,
\end{equation}
with the following version of the extended Kraft inequality:

\noindent
{\em Lemma 4.}
Let $(\phi_n,\psi_n)$ be a rate-$R$ fixed-rate code.
Then, for every $\alpha\ge 0$ and $\beta\ge 0$,
\begin{equation}
Z_{\mbox{\tiny fr}}^n(\alpha,\beta)\le
2^{n(1-\alpha)R}\cdot\left[\int_{\calU}2^{-\beta\rho(z)}\mbox{d}z\right]^n.
\end{equation}

\section{Lower Bounds}
\label{lowerbounds}

To fix ideas, we first demonstrate how the classical SLB is obtained from
Lemma 1. Consider an arbitrary encoder-decoder pair, $(\phi_n,\psi_n)$, in which
the reproduction vector, $v^n=\psi_n(\phi_n(u^n))$ is losslessly compressed by
a UD code. Then, following Lemma 1, we have the following chain of
inequalities:
\begin{eqnarray}
\label{lb}
\left[\int_{\calU}
2^{-\beta\rho(z)}\mbox{d}z\right]^n&\gea&Z_n(\alpha,\beta)\nonumber\\
&=&\int_{\calU^n}\exp_2\{-\alpha
L[\phi_n(u^n)]-\beta\rho(u^n-\psi_n(\phi_n(u^n)))\}\mbox{d}u^n\nonumber\\
&=&\int_{\calU^n}P(u^n)\exp_2\{-\alpha
L[\phi_n(u^n)]-\beta\rho(u^n-\psi_n(\phi_n(u^n)))-\log P(u^n)\}\mbox{d}u^n\nonumber\\
&=&\bE\left\{\exp_2\{-\alpha
L[\phi_n(U^n)]-\beta\rho(U^n-\psi_n(\phi_n(U^n)))-\log P(U^n)\}\right\}\nonumber\\
&\geb&\exp_2\left[-\alpha
\bE\{L[\phi_n(U^n)]\}-\beta\bE\{\rho(U^n-\psi_n(\phi_n(U^n)))\}-\bE\{\log
P(U^n)\}\right]\nonumber\\
&=&\exp_2\left[-\alpha
\bE\{L[\phi_n(U^n)]\}-\beta\bE\{\rho(U^n-\psi_n(\phi_n(U^n)))\}+h(U^n)
\right],
\end{eqnarray}
where (a) stems from Lemma 1 and (b) is due to Jensen's inequality applied to
the convex function, $f(t)=2^t$. It follows that
\begin{equation}
\label{lag1}
\alpha\bE\{L[\phi_n(U^n)]\}+\beta\bE\{\rho(U^n-\psi_n(\phi_n(U^n)))\}\ge
h(U^n)-n\log\left[\int_{\calU}2^{-\beta\rho(z)}\mbox{d}z\right].
\end{equation}
Since this holds true for every $\alpha>1$ while the right-hand side is
independent of $\alpha$, we may take the infimum of the left-hand side in the
range $\alpha> 1$ to obtain, after normalization by $n$,
\begin{equation}
\label{lag2}
\frac{\bE\{L[\phi_n(U^n)]\}}{n}+\beta\cdot\frac{\bE\{\rho(U^n-\psi_n(\phi_n(U^n)))\}}{n}\ge
\frac{h(U^n)}{n}-\log\left[\int_{\calU}2^{-\beta\rho(z)}\mbox{d}z\right].
\end{equation}
At this point, there are several possible perspectives that can be adopted
regarding eq.\ (\ref{lag2}). The first is, of course, to view this as a lower
bound to the Lagrangian of rate and distortion. The second is to impose an
expected distortion constraint, $\bE\{\rho(U^n-\psi_n(\phi_n(U^n)))\}\le nD$,
and then, after normalization by $n$, (\ref{lag2}) would yield a lower bound
to the expected rate according to
\begin{equation}
\frac{\bE\{L[\phi_n(U^n)]\}}{n}\ge
\frac{h(U^n)}{n}-
\log\left[\int_{\calU}2^{-\beta\rho(z)}\mbox{d}z\right]-\beta D.
\end{equation}
Since this holds true for every $\beta\ge 0$ while the left-hand side is
independent of $\beta$, we may maximize the right-hand side over $\beta\ge 0$,
to obtain
\begin{equation}
\frac{\bE\{L[\phi_n(U^n)]\}}{n}\ge \frac{h(U^n)}{n}-\inf_{\beta\ge
0}\left\{\log\left[\int_{\calU}2^{-\beta\rho(z)}\mbox{d}z\right]+\beta
D\right\}=\frac{h(U^n)}{n}-\Phi(D),
\end{equation}
thus recovering the classical SLB. 
In case of multiple distortion constraints, say,
\begin{equation}
\bE\{\rho_j(U^n-\psi_n(\phi_n(U^n)))\}\le nD_j,~~~~~j=1,2,\ldots,k,
\end{equation}
the above rate lower bound continues to apply, provided that $\beta$, $D$
and $\rho(z)$ are redefined as $k$-dimensional vectors, $\beta=(\beta_1,\ldots,\beta_k)$,
$D=(D_1,\ldots,D_k)$, and $\rho(z)=(\rho_1(z),\ldots,\rho_k(z))$, and
accordingly, $\beta D$
and $\beta\rho(z)$ are understood to be inner products. The infimum over
$\beta$ is then defined across $[0,\infty)^k$.

Returning to eq.\ (\ref{lag2}) and to a single distortion criterion, we may alternatively apply a rate
constraint $\bE\{L[\phi_n(U^n)]\}\le nR$ and then obtain a lower bound to the
expected distortion:
\begin{eqnarray}
\frac{\bE\{\rho(U^n-\psi_n(\phi_n(U^n)))\}}{n}
&\ge&\sup_{\beta\ge
0}\frac{1}{\beta}\left(\frac{h(U^n)}{n}-\log\left[\int_{\calU}2^{-\beta\rho(z)}\mbox{d}z\right]-R\right)\nonumber\\
&=&\sup_{\gamma\ge
0}\gamma\left(\frac{h(U^n)}{n}-\log\left[\int_{\calU}2^{-\rho(z)/\gamma}\mbox{d}z\right]-R\right),
\end{eqnarray}
which is the distortion-rate counterpart of the SLB.

Having established the classical SLB for general variable-rate codes with UD
lossless compression of the reproduction vectors, we now carry on to derive
some refinements and extensions associated with the other types of codes that
we mentioned. The idea would be to apply the same chain of inequalities as in
(\ref{lb}), but to invoke Lemma 2 or Lemma 3, or Lemma 4, according to the relevant
class of codes, instead of Lemma 1. We next implement this plan for
one-to-one codes and for $D$-semifaithful codes. The same methodology can be
applied to fixed-rate codes, but will not delve into this here.

\subsection{One-to-One Codes}
\label{121}

For one-to-one codes, we repeat the same derivation as in (\ref{lb}) by invoking Lemma
2 instead of Lemma 1. This yields the following modified version of eq.\ (\ref{lag1}):
\begin{equation}
\alpha\bE\{L[\phi_n(U^n)]\}+\beta\bE\{\rho(U^n-\psi_n(\phi_n(U^n)))\}\ge
h(U^n)-n\log\left[\int_{\calU}2^{-\beta\rho(z)}\mbox{d}z\right]+\log(2^{\alpha-1}-1).
\end{equation}
Applying the the average distortion constraint, optimizing over $\beta$, and normalizing by $n$, we
obtain
\begin{equation}
\frac{\bE\{L[\phi_n(U^n)]\}}{n}\ge\sup_{\alpha\ge 1}\left\{
\frac{h(U^n)/n-\Phi(D)}{\alpha}
+\frac{\log(2^{\alpha-1}-1)}{\alpha n}\right\}.
\end{equation}
The maximization of the right-hand side w.r.t.\ $\alpha$ does not seem to lend
itself to a closed form expression,
but by  selecting $\alpha=\alpha_n=1+c\frac{\log n}{n}$ ($c$ being an arbitrary
positive constant), it is readily seen that
the resulting lower bound becomes
\begin{equation}
\frac{\bE\{L[\phi_n(U^n)]\}}{n}\ge\frac{h(U^n)}{n}-\Phi(D)-O\left(\frac{\log
n}{n}\right),
\end{equation}
which is in agreement with the subtraction of an $O((\log
n)/n)$ below the entropy for lossless one-to-one
codes \cite{Rissanen82}. This reduction of $O((\log n)/n)$ is
due to the fact that the class of one-to-one codes is broader than the class
of UD codes, and therefore, the former codes are potentially more capable for
a given finite $n$, albeit
the difference fades away as $n$ grows. 

\subsection{$D$-Semifaithful Codes}
\label{Dsf}

For $D$-semifaithful codes with UD correspondence between the compressed
bit-stream and the reproduction, we obtain from Lemma 3, combined with a
derivation like (\ref{lb}):
\begin{equation}
\label{logvolslb}
\bE\{L[\phi_n(U^n)]\}\ge
h(U^n)-\log\mbox{Vol}\{\calS_n(D)\},
\end{equation}
which is line with the results of \cite{me95}, and the remaining issue becomes
the assessment of the log-volume of $\calS_n(D)$, or an evaluation of a good
upper bound to this quantity. One way to proceed is to
apply a Chernoff bound, as was
suggested by Campbell \cite{Campbell73}. For $\calU=\reals$, this amounts to the following
derivation:
\begin{eqnarray}
\label{chernoff}
\log\mbox{Vol}\{\calS_n(D)\}&=&\log\mbox{Vol}\{\bz:~\rho(\bz)\le
nD\}\nonumber\\
&=&\log\left[\int_{\reals^n}\calI\{\rho(\bz)\le
nD\}\mbox{d}\bz\right]\nonumber\\
&\le&\log\left(\inf_{\beta\ge
0}\int_{\reals^n}\exp_2\{\beta[nD-\rho(\bz)\}\mbox{d}\bz\right)\nonumber\\
&=&n\cdot\inf_{\beta\ge 0}\left\{\beta
D+\log\left(\int_{\reals}2^{-\beta\rho(z)}\mbox{d}z\right)\right\}\nonumber\\
&=&n\Phi(D),
\end{eqnarray}
and we are back to the ordinary SLB,
\begin{equation}
\frac{\bE\{L[\phi_n(U^n)]\}}{n}\ge \frac{h(U^n)}{n}-\Phi(D),
\end{equation}
exactly as we had for UD lossless compression of the reproduction data, and
once again, this derivation extends straightforwardly to the case of multiple
simultaneous distortion
constraints by considering $\beta$, $D$, and $\rho(\cdot)$ to be vectors
rather than scalars, as mentioned before.

However, since the class of $D$-semifaithful codes is narrower (and hence
more limited to a certain extent) than the class of
codes that merely comply with an average distortion constraint, it is
conceivable to
expect a somewhat tighter (larger) lower bound. Indeed, this is turns out to
be the case if 
the log-volume of $\calS_n(D)$ is estimated using a more sophisticated analysis
tool, namely, the saddle-point method (see, e.g., Chapter 5 in
\cite{deBruijn81} and Chapter 3 in \cite{MW25} as well as its extension to
the multivariate case \cite{Neuschel14}).

To apply the saddle-point method, the idea is to represent the indicator
function at the integrand of the second line in (\ref{chernoff}) as
$\calI\{\rho(\bz)\le nD\}=u(nD-\rho(\bz))$, where $u(t)$ is the unit step
function, defined as 
\begin{equation}
u(t)=\left\{\begin{array}{ll}
0 & t<0\\
1 & t\ge 0\end{array}\right.
\end{equation}
which in turn is represented as the inverse Laplace transform of the complex
function $U(s)=\frac{1}{s}$, i.e.,
\begin{equation}
u(t)=\frac{1}{2\pi
i}\int_{\mbox{\small Re}\{s\}=c}\frac{e^{st}\mbox{d}s}{s},
\end{equation}
where $i\dfn\sqrt{-1}$ and $c$ is an arbitrary positive real. It follows that 
\begin{eqnarray}
\mbox{Vol}\{S_n(D)\}&=&\int_{\reals^n}u(nD-\rho(\bz))\mbox{d}\bz\nonumber\\
&=&\int_{\reals^n}\frac{\mbox{d}\bz}{2\pi
i}\int_{\mbox{\small Re}\{s\}=c}\frac{e^{s(nD-\rho(\bz))}\mbox{d}s}{s}\nonumber\\
&=&\frac{1}{2\pi i}
\int_{\mbox{\small Re}\{s\}=c}\frac{e^{snD}\mbox{d}s}{s}\int_{\reals^n}e^{-s\rho(\bz)}\mbox{d}\bz\nonumber\\
&=&\frac{1}{2\pi i}
\int_{\mbox{\small Re}\{s\}=c}\frac{e^{snD}\mbox{d}s}{s}\int_{\reals^n}\exp\left\{-s\sum_{t=1}^n\rho(z_t)\right\}\mbox{d}\bz\nonumber\\
&=&\frac{1}{2\pi i}
\int_{\mbox{\small Re}\{s\}=c}\frac{e^{snD}\mbox{d}s}{s}\left[\int_{\reals}e^{-s\rho(z)}\mbox{d}z\right]^n\nonumber\\
&=&\frac{1}{2\pi i}
\int_{\mbox{\small Re}\{s\}=c}\frac{\mbox{d}s}{s}\exp\left\{n\left[sD+\ln\left(\int_{\reals}e^{-s\rho(z)}\mbox{d}z\right)\right]\right\}.
\end{eqnarray}
This path integral in the complex plane complies with the general
form $\int_A^B g(s)e^{nf(s)}\mbox{d}s$, which under certain regularity
conditions, can be approximated for large $n$ (see, e.g., eq.\ (5.7.2) in \cite{deBruijn81})
according to:
\begin{equation}
\int_A^B
g(s)e^{nf(s)}\mbox{d}s=e^{i\theta}\sqrt{\frac{2\pi}{n|f''(s_\star)|}}\cdot
g(s_\star)e^{nf(s_\star)}\cdot\left[1+O\left(\frac{1}{n}\right)\right],
\end{equation}
provided that the functions $f$ and $g$ are independent of $n$ and analytic within some
connected region $\calD$ that includes $A$ and $B$ (which are also independent of
$n$). Here,
$s_\star\in\calD$ is a saddle-point, i.e., a point where $f'(s_\star)=0$,
$f''(s_\star)>0$ and $g(s_\star)\ne 0$. The angle $\theta$ is called the axis and is
given by $\theta=(\pi-\mbox{arg}\{f''(s_\star)\})/2$. In our case, 
$f(s)=sD+\ln\left[\int_\reals e^{-s\rho(z)}\mbox{d}z\right]$,
$g(s)=\frac{1}{s}$, $\theta=\frac{\pi}{2}$, and $s_\star$ is the point at which $f'$
vanishes, which is assumed strictly positive. But this point of zero
derivative is also the point
that minimizes the convex function $f$ across the positive reals.
It follows then that $\mbox{Vol}\{S_n(D)\}$ can be approximated as follows:
\begin{equation}
\mbox{Vol}\{S_n(D)\}=\frac{1}{s_\star\sqrt{2\pi
n|f''(s_\star)|}}\cdot\exp\left\{n\left(s_\star D+\ln\left[\int_\reals
e^{-s_\star\rho(z)}\mbox{d}z\right]\right)\right\}\cdot\left[1+O\left(\frac{1}{n}\right)\right].
\end{equation}
As for the exponential factor, observe that
\begin{eqnarray}
& &\exp\left\{n\left(s_\star D+\ln\left[\int_\reals
e^{-s_\star\rho(z)}\mbox{d}z\right]\right)\right\}\nonumber\\
&=&\exp\left\{n\cdot\inf_{s\ge 0}\left(sD+\ln\left[\int_\reals
e^{-s\rho(z)}\mbox{d}z\right]\right)\right\}\nonumber\\
&=&\exp_2\left\{n(\log_2 e)\cdot\inf_{s\ge 0}\left(sD+\frac{1}{\log_2 e}\cdot\log_2\left[\int_\reals
2^{-s\rho(z)\log_2 e}\mbox{d}z\right]\right)\right\}\nonumber\\
&=&\exp_2\left\{n\inf_{s\ge 0}\left(sD\log_2 e+\log_2\left[\int_\reals
2^{-s\rho(z)\log_2 e}\mbox{d}z\right]\right)\right\}\nonumber\\
&=&\exp_2\left\{n\inf_{\beta\ge 0}\left(\beta D+\log_2\left[\int_\reals
2^{-\beta\rho(z)}\mbox{d}z\right]\right)\right\}\nonumber\\
&=&2^{n\Phi(D)},
\end{eqnarray}
and so,
\begin{equation}
\mbox{Vol}\{S_n(D)\}=\frac{2^{n\Phi(D)}}{s_\star\sqrt{2\pi
n|f''(s_\star)|}}\cdot
\left[1+O\left(\frac{1}{n}\right)\right],
\end{equation}
which yields
\begin{equation}
\frac{\log \mbox{Vol}\{S_n(D)\}}{n}=\Phi(D)-\frac{\log
n}{2n}-O\left(\frac{1}{n}\right),
\end{equation}
and then, following eq.\ (\ref{logvolslb}), we end up with
\begin{equation}
\frac{\bE\{L[\phi_n(U^n)]\}}{n}\ge \frac{h(U^n)}{n}-\Phi(D)+\frac{\log
n}{2n}+O\left(\frac{1}{n}\right).
\end{equation}
We therefore observe that the SLB for $D$-semifaithful codes is given by the
ordinary SLB plus 
redundancy whose leading term is $\frac{\log n}{2n}$.
This is in agreement with findings of \cite{me95},
where the derivations corresponded to special cases for which simple geometric and/or
combinatorial considerations facilitated the
accurate characterization of $\log\mbox{Vol}\{\calS_n(D)\}$, but here the conclusion
is more general. Note that
we could have been even more precise and specify also
the $O\left(\frac{1}{n}\right)$ term to be
$\frac{1}{n}\cdot\log[s_\star\sqrt{2\pi
|f''(s_\star)|}]+O\left(\frac{1}{n^2}\right)$, but this is, of course, less important.

When $k$ simultaneous distortion constraints are imposed pointwise, i.e.,
\begin{equation}
\max_{u^n\in\reals^n}\rho_j(u^n-\psi_n(\phi_n(u^n)))\le nD_j,~~~j=1,2,\ldots,k,
\end{equation}
then using the multivariate saddle-point method \cite{Neuschel14},
the above derivation extends with the vector version of the definition of
$\Phi(D)$ replacing the scalar one, and the pre-exponential factor
of the saddle-point approximation becomes 
$$\frac{(2\pi/n)^{k/2}}{Q\sqrt{\mbox{det}\{\mbox{Hess}\{f\}|_{s_\star}\}}},$$
where $Q$ is the product of the components of the $k$-dimensional vector
$s_\star$ -- the point at which $\nabla f(s)=0$, where $f(s)=s\cdot D+\ln\left[\int_\reals
e^{-s\cdot\rho(z)}\mbox{d}z\right]$ is defined such that $s$, $D$ and $\rho(\cdot)$ are
$k$-dimensional vectors with $s\cdot D$ and $s\cdot \rho(z)$ being inner products, as was defined
before. Here, $\mbox{det}\{\mbox{Hess}\{f\}|_{s_\star}$ is 
the determinant of the Hessian of $f$, computed at $s=s_\star$. It is assumed,
of course, that the $k\times k$ matrix $\mbox{Hess}\{f\}|_{s_\star}$ is
non-singular, otherwise there might be redundant (inactive) constraints, which should be
removed from the calculation.
In this case, the redundancy on top of the ordinary SLB is $\frac{k'\log
n}{2n}+O\left(\frac{1}{n}\right)$, i.e.,
\begin{equation}
\frac{\bE\{L[\phi_n(U^n]\}}{n}\ge\frac{h(U^n)}{n}-\Phi(D)+\frac{k'\log
n}{2n}+O\left(\frac{1}{n}\right),
\end{equation}
where $k'\le k$ is the effective dimension
after the possible removal of redundant constraints.

For example, if $k=2$, $\rho_1(z)=|z|$, and $\rho_2(z)=z^2$, then obviously,
$\left(\frac{1}{n}\sum_{t=1}^n|z_t|\right)^2$ cannot exceed
$\frac{1}{n}\sum_{t=1}^nz_t^2$, and so, if $D_1> \sqrt{D_2}$, the constraint 
$\sum_{t=1}^n|z_t|\le nD_1$ is inactive (and hence removable) in the presence of the constraint
$\sum_{t=1}^nz_t^2\le nD_2$. In this case, although $k=2$, the effective
dimension is $k'=1$. Another obvious example of a superfluous distortion constraint
occurs when there is a linear dependence. For instance, let $k=3$ and
$\rho_3(z)=a\rho_1(z)+b\rho_2(z)$, where $a$ and $b$ are fixed positive
reals. Then whenever $D_3\ge aD_1+bD_2$, the third distortion constraint is
redundant and then $k'=2$ (or even less, if there are additional superfluous
constraints). More generally,
inactive constraints can be identified as
those whose corresponding components of $s_\star$ vanish.

Another type of situations where the effective dimension $k'$ is smaller than
the formal dimension, $k$, occurs when the dependence of $f(s)$ on some of the
components of $s$ disappears in the first place.
Consider, for example, the case where $k=2$, $\rho_1(z)=z^2$, and the
other distortion constraint is $\max_{1\le t\le n}|z_t|\le A$, in other words, we impose
both an average quadratic constraint and a peak-limited distortion constraint
with the motivation of avoiding large spikes in the reconstruction error
signal. The peak-limited distortion constraint can be represented as an additive distortion constraint if we
select $\rho_2(z)=W(|z|-A)$, where $W(\cdot)$ is the IWF defined in (\ref{iwf})
and the value of $D_2$ can then be selected to be an arbitrary finite, non-negative, real, say $D_2=0$. Here
again $k'=1$, but as said, this time, it is because
the function $f(s)$ depends only on one component of the vector $s=(s_1,s_2)$
to begin with. To see this, observe that
\begin{eqnarray}
f(s)&=&sD+\ln\left[\int_\reals
e^{-s\rho(z)}\mbox{d}z\right]\nonumber\\
&=&s_1D_1+s_2D_2+\ln\left[\int_\reals
e^{-s_1\rho_1(z)+s_2\rho_2(z)}\mbox{d}z\right]\nonumber\\
&=&s_1D_1+s_2\cdot 0+\ln\left[\int_\reals
e^{-s_1z^2+s_2W(|z|-A)}\mbox{d}z\right]\nonumber\\
&=&s_1D_1+\ln\left[\int_{-A}^A
e^{-s_1z^2}\mbox{d}z\right]\nonumber\\
&=&s_1D_1+\ln\left(\sqrt{\frac{\pi}{s_1}}\cdot[1-2Q(A\sqrt{2s_1})]\right)\nonumber\\
&=&s_1D_1+\frac{1}{2}\ln\left(\frac{\pi}{s_1}\right)+\ln[1-2Q(A\sqrt{2s_1})],
\end{eqnarray}
where $Q(\cdot)$ is the well-known $Q$-function, defined as
\begin{equation}
Q(t)=\int_t^\infty\frac{e^{-x^2/2}\mbox{d}x}{\sqrt{2\pi}}.
\end{equation}
Here, although formally there are $k=2$ distortion constraints, the saddle-point
integration is over one complex variable only, and so, the redundancy is
$\frac{\log n}{2n}+O\left(\frac{1}{n}\right)$ on top of the ordinary SLB,
$\frac{h(U^n)}{n}+\Phi(D)$.

\section{Sliding-Window Distortion Constraints}
\label{slidingwindow}

Consider a situation where one wishes to control not only the `intensity' of
the reconstruction error signal, $\sum_{t=1}^n\rho(z_i)$, but possibly to shape
also its `continuity' or its `smoothness' by imposing additional limitations,
say, on the
empirical autocorrelations of the error signal, $z^n=(z_1,z_2,\ldots,z_n)$, in
order to suppress, for example, high frequencies, which might be disturbing for the human
eye (in case of images and video streams) or the human ear (in case of audio
signals). For instance, one might wish to constrain the error signal to obey
the first lag autocorrelation constraint, $\frac{1}{n}\sum_{t=2}^nz_tz_{t-1}\ge 0.95$
(and perhaps also further lags), in addition to the ordinary
mean-square error constraint, say, $\frac{1}{n}\sum_{t=1}^nz_t^2\le 0.01$.
More generally, consider the case where we impose $k$ additive constraints
pertaining to sliding-window functions, of the form,
\begin{equation}
\sum_{t=m}^n\rho_j(z_{t-m+1}^t)\le nD_j,~~~~~j=1,2,\ldots,k,
\end{equation}
where $m$ is a positive integer that designates the size of the sliding
window. For $m=1$, we are back to ordinary additive distortion constraints,
where the single-letter distortion function $\rho_j$ operates on single
symbols separately. If the sizes of the sliding window are different for the $k$ various
constraints, which take $m$ to be the largest one. Accordingly, in the above
example where $k=2$ and the constraints are $\frac{1}{n}\sum_{t=1}^nz_t^2\le 0.01$
and $\frac{1}{n}\sum_{t=2}^nz_tz_{t-1}\ge 0.95$, we have $\rho_1(z)=z^2$,
$D_1=0.01$, $\rho_2(z_1,z_2)=-z_1\cdot z_2$, and $D_2=-0.95$. In this case,
the sliding-window size is $m=2$, and the minus signs in the correlation
constraint are due to the reversal of the direction of the inequality in
$\frac{1}{n}\sum_{t=2}^nz_tz_{t-1}\ge 0.95$, as opposed to the direction
of inequalities in our distortion
constraints in general.

How does the SLB extend to the case of sliding-window constraints of this
type? In this section, we assume that $m$ is fixed while $n$ tends to
infinity, and we focus merely on the main terms of the resulting SLB,
disregarding the redundancy terms. 

A straightforward extension of Lemma 1 and the subsequent derivation in
Section \ref{lowerbounds} so as to apply to a set of $k$ sliding-window
distortion constraints with
window size $m$, yields the lower bound
\begin{equation}
\bE\{L[\phi(U^n)]\}\ge h(U^n)-\inf_{\beta\in[0,\infty)^k}
\left(\log\left[\int_{\calU^n}\exp_2\left\{-\beta\cdot\sum_{t=m}^n\rho(z_{t-m+1}^t)\right\}\mbox{d}z^n\right]+n\beta\cdot
D\right),
\end{equation}
where $\beta=(\beta_1,\ldots,\beta_k)$,
$D=(D_1,\ldots,D_k)$,
$\rho(z_{t-m+1}^t)=(\rho_1(z_{t-m+1}^t),\ldots,\rho_k(z_{t-m+1}^t))$, and
the dot operations are understood as inner products, as defined earlier.

The multi-dimensional integral 
\begin{equation}
\int_{\calU^n}\exp_2\left\{-\beta\cdot\sum_{t=m}^n\rho(z_{t-m+1}^t)\right\}\mbox{d}z^n
=\int_{\calU^n}\prod_{t=m}^n\exp_2\left\{-\beta\cdot\rho(z_{t-m+1}^t)\right\}\mbox{d}z^n
\end{equation}
can be viewed as being obtained by iterated applications of a sliding-window
integral operator whose kernel is given by
$K_\beta(z_{t-m+1}^t)=\exp_2\left\{-\beta\cdot\rho(z_{t-m+1}^t)\right\}$
and the integration is over one component of $z^n$ at a time. Under certain
regularity conditions, the value of this multidimensional integral grows
exponentially with an exponential order of $[\lambda(\beta)]^n$, where
$\lambda(\beta)$ is the spectral radius of the operator kernel, namely, the
dominant eigenvalue, and then the asymptotic form of the SLB becomes
\begin{equation}
\liminf_{n\to\infty}\frac{\bE\{L[\phi_n(U^n)]\}}{n}\ge
\bar{h}(U^\infty)-\inf_{\beta\in[0,\infty)^k}[\log\lambda(\beta)+\beta D].
\end{equation}
There are several equivalent formulas for calculating the spectral radius,
$\lambda(\beta)$, of the
of a sliding-window kernel $K_\beta(\cdot)$ for a given $\beta$, for example, the Collatz-Wielandt
formula \cite{Collatz42}, \cite{Wielandt50}, and the Donsker-Varadhan formula
\cite{DV83} -- for details, see, e.g., Subsection 3.4 of \cite{MS25}. In
certain special cases, such as those that involve a symmetric kernel for $m=2$,
$K_\beta(z,z')$, the
Rayleigh quotient formula
\begin{equation}
\lambda(\beta)=\sup_g\frac{\int_{U^2}g(z)K_\beta(z,z')g(z')\mbox{d}z\mbox{d}z'}{\int_{\calU}g^2(z)\mbox{d}z}.
\end{equation}
can also be used. Clearly, in the finite-alphabet case, the operator kernel reduces to
a finite-dimensional matrix and then the spectral radius is simply the
Perron-Frobenius eigenvalue (see, e.g., Theorem 8.2.11 of \cite{HJ85}).

As a simple example, consider the case $k=m=2$, $\rho_1(z)=z^2$, $D_1=D$,
$\rho_2(z,z')=-z\cdot z'$, and $D_2=-\theta$, where $D>0$ and
$\theta\in[0,1)$ are given parameters. Then, following the analogous
derivation of Example 4 in \cite{MS25}, we find that the resulting SLB is
given by
\begin{equation}
\liminf_{n\to\infty}
\frac{\bE\{L[\phi_n(U^n)]\}}{n}\ge\bar{h}(U^\infty)-\frac{1}{2}\log[2\pi e
D(1-\theta^2)],
\end{equation}
or, in words, the rate penalty due to the additional autocorrelation constraint
is $\frac{1}{2}\log\frac{1}{1-\theta^2}$ on top of the SLB pertaining to the
ordinary quadratic distortion constraint,
$\frac{h(U^n)}{n}-\frac{1}{2}\log(2\pi eD)$.

\section{Individual Sequences and Finite-State Encoders}
\label{indivseq}

We conclude this article by deriving an individual-sequence counterpart of the
SLB for finite-alphabet, deterministic sequences, which is based on a variation of Ziv and
Lempel's generalized Kraft inequality (see Lemma 2 of \cite{ZL78}).

Generally speaking, our model for lossy compression of individual sequences is
based on the following simple structure. Each source block of length $m$,
$u^m\in\calU^m$ ($m$ -- positive integer)
is first mapped by an arbitrary reproduction encoder (or vector
quantizer) into a reproduction vector, 
\begin{equation}
\label{vq}
v^m=q(u^m)\in\calV_m\subseteq\calU^m
\end{equation}
and then, the concatenation of the resulting $m$-vectors, $\{v^m\}$, forming
the sequence $v_1,v_2,\ldots$, is compressed losslessly by a
finite-state encoder following the model of \cite{ZL78}. 

To keep this article self-contained, we begin with some basic background on
lossless compression of individual sequences by finite-state encoders and the
1978 Lempel-Ziv (LZ78) algorithm. Readers familiar with this background may
safely skip Subsection \ref{indivbg} and move on
directly to Subsection \ref{indivslb}.

\subsection{Background}
\label{indivbg}

Following the model of \cite{ZL78}, consider a setting for lossless
compression of $v^n$ on the basis of finite-state (FS) encoders. 
An FS encoder is
defined by the set
$E=(\calU,\calY,\Sigma,f,g)$,
where: $\calU$ is the finite alphabet of each symbol, $v_i$, which is the same as the
source alphabet, and whose size is $r$;
$\calY$ is a finite collection of binary, variable-length strings, which is
allowed to consist of empty string $\lambda$ (whose length is zero);
$\Sigma$ is a set of $s$ states of the encoder;
$f: \Sigma \times \calU \to \calY$ is the output function, and
$g: \Sigma \times \calU \to \Sigma$ is the next-state function.
Given an infinite input reproduction vector (obtained by concatenating
infinitely many output vectors from the reproduction encoder), $\bv = (v_1, v_2, \ldots)$ with $v_i \in
\calU$, $i=1,2,\ldots$, the FS encoder $E$ produces an
infinite output sequence, $\by = (y_1, y_2, \ldots)$ with $y_i \in \calY$,
henceforth referred to as the compressed bit-stream, while passing through a sequence of 
states $\bsigma = (\sigma_1, \sigma_2, \ldots)$ with $\sigma_i \in \Sigma$.
The encoder is governed recursively by the equations:
\begin{eqnarray}
y_i &=& f(\sigma_i, v_i), \label{yi} \\
\sigma_{i+1}&=&g(\sigma_i,v_i), \label{nextstate}
\end{eqnarray}
for $i = 1, 2, \ldots$, with a fixed initial state $\sigma_1 = \sigma_\star\in\calZ$.
If at any step $y_i = \lambda$, this is referred to as idling as no output is
generated, but only the state is kept updated in response to the input.

An encoder with $s$ states, henceforth called an $s$-state
encoder, is one for which $|\Sigma| = s$.
For the sake of simplicity, we adopt a few notation conventions from \cite{ZL78}:
Given a segment of input symbols $v_i^j$ with $i \le j$ and an initial state
$\sigma_i$, we use $f(\sigma_i, v_i^j)$ to denote the corresponding output
segment $y_i^j$ produced by $E$.
Similarly, $g(\sigma_i, v_i^j)$ will denote the final state $\sigma_{j+1}$ after processing the
inputs $v_i^j$, beginning from state $\sigma_i$.

An FS encoder $E$ is called information lossless (IL) if,
given any initial state $\sigma_i \in \Sigma$, any positive integer $n$, and any
input string, $v_i^{i+n}$, the set
$(\sigma_i,f(\sigma_i,v_i^{i+n}),g(\sigma_i,v_i^{i+n}))$
uniquely determines the corresponding input string $v_i^{i+n}$. 

The incremental parsing process used by the LZ78 algorithm is a sequential
procedure for processing a finite-alphabet input $u^n$.
At each step of this process, one determines the shortest string that has not
yet occurred as a complete phrase
in the current parsed set with the possible exception of the last phrase,
which might be incomplete.
For example, applying this parsing method to the sequence
$$u^{15}=\mbox{011010011000100}$$
yields
$$\mbox{0,1,10,100,11,00,01,00}.$$
Let us denote by $c(u^n)$ the total number of distinct phrases generated by this procedure
(here, $c(u^{15}) = 8$). In addition, let $LZ(u^n)$ represent the length
in bits of the binary string produced by the LZ78 encoding of $u^n$.
By Theorem 2 in \cite{ZL78}, the following inequality holds:
\begin{equation}
\label{lz-clogc}
LZ(u^n)\le[c(u^n)+1]\log\{2r[c(u^n)+1]\}
\end{equation}
which can easily be shown to be further upper bounded by
\begin{equation}
\label{epsilon1}
LZ(u^n)\le c(u^n)\log c(u^n)+n\cdot\epsilon_1(n),
\end{equation}
where $\epsilon_1(n)$ tends to zero uniformly as $n\to\infty$.
In words, the LZ78 code length for $v^n$ is upper bounded by an expression whose
leading term is $c(u^n)\log c(u^n)$. 
We shall refer to the quantity
$c(u^n)\log c(u^n)$ as
the unnormalized LZ complexity of $u^n$, to distinguish from 
the normalized LZ complexity defined as
$\frac{c(u^n)\log
c(u^n)}{n}$,
which means the per-symbol LZ complexity.

\subsection{SLB for Individual Sequences}
\label{indivslb}

The following lemma provides a variation of the generalized Kraft inequality
of \cite{ZL78}.\\

\noindent
{\em Lemma 5.}
For every IL FS encoder with $s$ states, every $\alpha>1$, every $\beta\ge 0$,
and every positive integer $\ell$ which is an integer multiple of $m$,
\begin{equation}
\sum_{\sigma\in\Sigma}\sum_{w^\ell\in\calU^\ell}\exp_2\left\{-\alpha
L[f(\sigma,q(w^\ell))]-\beta\rho(w^\ell-q(w^\ell))\right\}\le
\frac{s^2\left[\sum_{z\in\calU}2^{-\beta\rho(z)}\right]^\ell}{2^{\alpha-1}-1},
\end{equation}
where $q(w^\ell)\equiv q(w_1^m,w_{m+1}^{2m},\ldots,w_{\ell-m+1}^\ell)\dfn
(q(w_1^m),q(w_{m+1}^{2m}),\ldots,q(w_{\ell-m+1}^\ell))$, the latter being
defined as in (\ref{vq}) for vectors in $\calU^\ell$.\\

\noindent
{\em Proof.} 
From the postulated IL property, it follows that given $\sigma\in\Sigma$, there cannot be
more than $s2^k$ distinct vectors, $\{v^\ell\}$, such that
$L[f(\sigma,v^\ell)]=k$, for every positive integer $k$. Therefore,
\begin{eqnarray}
& &\sum_{\sigma\in\Sigma}\sum_{w^\ell\in\calU^\ell}\exp_2\left\{-\alpha
L[f(\sigma,q(w^\ell))]-\beta\rho(w^\ell-q(w^\ell))\right\}\nonumber\\
&=&\sum_{\sigma\in\Sigma}\sum_{k\ge 1}
\sum_{\{v^\ell:~L[f(\sigma,v^\ell)]=k\}}\sum_{\{w^\ell:~q(w^\ell)=v^\ell\}}\exp_2\left\{-\alpha k-
\beta\rho(w^\ell-q(w^\ell))\right\}\nonumber\\
&\le&\sum_{\sigma\in\Sigma}\sum_{k\ge 1} s\cdot 2^k\cdot 2^{-\alpha k}
\sum_{z^\ell\in\calU^\ell}2^{
-\beta\rho(z^\ell))}\nonumber\\
&=&s^2\cdot\left[\sum_{u\in\calU}2^{-\beta\rho(z)}\right]^\ell\cdot\sum_{k=1}^\infty
2^{-(\alpha-1)k}\nonumber\\
&=&\frac{s^2\left[\sum_{z\in\calU}2^{-\beta\rho(z)}\right]^\ell}{2^{\alpha-1}-1}.
\end{eqnarray}
This completes the proof of Lemma 5.\\

Now, let $\ell$ divide $n$. For a given FS IL encoder $E$, a given
$u^n\in\calU^n$, and the associated state sequence $\sigma^n\in\Sigma^n$
(generated from $u^n$ by $E$ using the next-state function $g$ recursively), consider the joint empirical distribution
\begin{equation}
\hat{P}(\sigma,w^\ell)=\frac{\ell}{n}\sum_{i=0}^{n/\ell-1}
\calI\{\sigma_{i\ell+1}=\sigma,~u_{i\ell+1}^{i\ell+\ell}=w^\ell\},~~~~\sigma\in\Sigma,~w^\ell\in\calU^\ell
\end{equation}
Then, according to Lemma 5,
\begin{eqnarray}
& &\frac{s^2\left[\sum_{z\in\calU}2^{-\beta\rho(z)}\right]^\ell}{2^{\alpha-1}-1}\nonumber\\
&\ge&\sum_{\sigma\in\Sigma}\sum_{w^\ell\in\calU^\ell}\exp_2\left\{-\alpha
L[f(\sigma,q(w^\ell))]-\beta\rho(w^\ell-q(w^\ell))\right\}\nonumber\\
&=&\sum_{\sigma\in\Sigma}\sum_{w^\ell\in\calU^\ell}\hat{P}(\sigma,w^\ell)\exp_2\left\{-\alpha
L[f(\sigma,q(w^\ell))]-\beta\rho(w^\ell-q(w^\ell))-\log\hat{P}(\sigma,w^\ell)\right\}\nonumber\\
&\ge&\exp_2\left\{-\alpha\sum_{\sigma\in\Sigma}\sum_{w^\ell\in\calU^\ell}\hat{P}(\sigma,w^\ell)
L[f(\sigma,q(w^\ell))]-\beta\sum_{w^\ell\in\calU^\ell}\hat{P}(w^\ell)\rho(w^\ell-q(w^\ell))+
\hat{H}(S,U^\ell)\right\}\nonumber\\
&=&\exp_2\left\{-\alpha\cdot\frac{\ell}{n}\sum_{i=0}^{n/\ell-1}L[f(\sigma_{i\ell+1},v_{i\ell+1}^{i\ell+\ell})]
-\beta\cdot\frac{\ell}{n}\sum_{i=0}^{n/\ell-1}\rho(u_{i\ell+1}^{i\ell+\ell}-q(u_{i\ell+1}^{i\ell+\ell}))+
\hat{H}(S,U^\ell)\right\}\nonumber\\
&=&\exp_2\left\{-\alpha\cdot\frac{\ell}{n}\sum_{t=1}^nL[f(\sigma_t,v_t)]
-\beta\cdot\frac{\ell}{n}\sum_{t=1}^n\rho(u_t-v_t)+
\hat{H}(S,U^\ell)\right\},
\end{eqnarray}
where $\hat{P}(w^\ell)=\sum_{\sigma\in\Sigma}\hat{P}(\sigma,w^\ell)$ and
$\hat{H}(S,U^\ell)$ is the joint entropy of the auxiliary random variables
$S\in\Sigma$
and $U^\ell\in\calU^\ell$, induced by the empirical joint distribution $\hat{P}$.
It follows then that
\begin{eqnarray}
& &\alpha\cdot\frac{1}{n}\sum_{t=1}^nL[f(\sigma_t,v_t)]
+\beta\cdot\frac{1}{n}\sum_{t=1}^n\rho(u_t-v_t)\nonumber\\
&\ge&\frac{\hat{H}(S,U^\ell)}{\ell}-\frac{\log(s^2)}{\ell}-\log\left[\sum_{z\in\calU}2^{-\beta\rho(z)}\right]+\frac{\log(2^{\alpha-1}-1)}{\ell}\nonumber\\
&\ge&\frac{\hat{H}(U^\ell)}{\ell}-
\log\left[\sum_{z\in\calU}2^{-\beta\rho(z)}\right]-\frac{2\log s}{\ell}+\frac{\log(2^{\alpha-1}-1)}{\ell}\nonumber\\
&\ge&\frac{c(u^n)\log c(u^n)}{n}-
\log\left[\sum_{z\in\calU}2^{-\beta\rho(z)}\right]-\Delta_n(\ell)-\frac{2\log
s}{\ell}+\frac{\log(2^{\alpha-1}-1)}{\ell},
\end{eqnarray}
where $\lim_{n\to\infty}\Delta_n(\ell)=\frac{1}{\ell}$ and the last step can
be found in eq.\ (12) of \cite{me25} as well as references therein. By taking
the limit of $\ell\to\infty$ (followed by the limit $n\to\infty$), the last
three terms can be made arbitrarily small, whereas the first two terms serve
as the individual-sequence counterpart of the right-hand side of eq.\
(\ref{lag2}). Moving the distortion term to the right-hand side and optimizing
over $\beta$, we obtain
\begin{eqnarray}
\alpha\cdot\frac{1}{n}\sum_{t=1}^nL[f(\sigma_t,v_t)]&\ge&
\frac{c(u^n)\log c(u^n)}{n}-\inf_{\beta\ge
0}\left\{\beta\cdot\frac{1}{n}\sum_{t=1}^n\rho(u_t-v_t)+
\log\left[\sum_{z\in\calU}2^{-\beta\rho(z)}\right]\right\}-\nonumber\\
& &\Delta_n(\ell)-\frac{2\log
s}{\ell}+\frac{\log(2^{\alpha-1}-1)}{\ell}\nonumber\\
&=&\frac{c(u^n)\log
c(u^n)}{n}-\Phi\left(\frac{1}{n}\sum_{t=1}^n\rho(u_t-v_t)\right)-\nonumber\\
& &\Delta_n(\ell)-\frac{2\log
s}{\ell}+\frac{\log(2^{\alpha-1}-1)}{\ell}.
\end{eqnarray}
Let $L_{\max}=\max_{\sigma,v}L[f(\sigma,v)]$,
Selecting $\alpha=1+\zeta_\ell$, where $\zeta_\ell$ tends to zero at
sub-exponential rate (say, $\zeta_\ell=1/\ell$), we have on the one hand,
the previous inequality, and on the other hand,
\begin{equation}
\alpha\cdot \frac{1}{n}\sum_{t=1}^nL[f(\sigma_t,v_t)]\le
\frac{1}{n}\sum_{t=1}^nL[f(\sigma_t,v_t)]+\zeta_\ell\cdot L_{\max},
\end{equation}
and so, our main result in this section is the following:
\begin{eqnarray}
\frac{1}{n}\sum_{t=1}^nL[f(\sigma_t,v_t)]&\ge&
\frac{c(u^n)\log
c(u^n)}{n}-\Phi\left(\frac{1}{n}\sum_{t=1}^n\rho(u_t-v_t)\right)-\nonumber\\
& &\Delta_n(\ell)-\frac{2\log
s}{\ell}+\frac{\log(2^{\zeta_\ell}-1)}{\ell}-\zeta_\ell\cdot
L_{\max}\nonumber\\
&\ge&\frac{c(u^n)\log
c(u^n)}{n}-\Phi\left(\frac{1}{n}\sum_{t=1}^n\rho(u_t-v_t)\right)-\nonumber\\
& &\Delta_n(\ell)-\frac{2\log
s}{\ell}+\frac{\log(\zeta_\ell\ln 2)}{\ell}-\zeta_\ell\cdot
L_{\max},
\end{eqnarray}
with the last four terms tending to zero in the limit of $\ell\to\infty$
followed by the limit $n\to\infty$.

To conclude, the individual-sequence counterpart of the SLB for
finite-alphabet source sequences is essentially of the same form as the
classical SLB of the probabilistic setting, except that the normalized entropy term is replaced by
the normalized LZ complexity of the sequence and the function $\Phi(\cdot)$ is
calculated at the point of the actual distortion,
$\frac{1}{n}\sum_{t=1}^n\rho(u_t-v_t)$.

\section*{Appendix }
\renewcommand{\theequation}{A.\arabic{equation}}
    \setcounter{equation}{0}

\noindent
{\em Proof of the second equality in eq.\ (\ref{equivalence}).}
Let us define the density function,
\begin{equation}
g(z)=\frac{2^{-\beta\rho(z)}}{\int_{\calU} 2^{-\beta\rho(z')}\mbox{d}z'}.
\end{equation}
Then,
\begin{eqnarray}
\inf_{\beta\ge 0}\left\{\log\left[\int_{\calU}
2^{-\beta\rho(z)}\mbox{d}z\right]+\beta D\right\}
&=&\inf_{\beta\ge 0}\left\{\sup_f[-D(f\|g)]+\log\left[\int_{\calU}
2^{-\beta\rho(z)}\mbox{d}z\right]+\beta D\right\}\nonumber\\
&=&\inf_{\beta\ge 0}\left\{\sup_f\left[\int_{\calU}\mbox{d}z
f(z)\log\left(\frac{2^{-\beta\rho(z)}}{f(z)}\right)\right]
+\beta D\right\}\nonumber\\
&=&\inf_{\beta\ge 0}\sup_f\left\{-\int_{\calU}\mbox{d}z
f(z)\log f(z)
+\beta\left[D-\int_{\calU} f(z)\rho(z)\mbox{d}z\right]\right\}\nonumber\\
&\eqa&\sup_f \inf_{\beta\ge 0}\left\{-\int_{\calU}\mbox{d}z
f(z)\log f(z)
+\beta\left[D-\int_{\calU} f(z)\rho(z)\mbox{d}z\right]\right\}\nonumber\\
&=&\sup_f \inf_{\beta\ge 0}\left\{h(Z)+\beta[D-\bE\{\rho(Z)\}]\right\}\nonumber\\
&=&\sup_f\left\{\begin{array}{ll}
-\infty & \bE\{\rho(Z)\}>D\nonumber\\
h(Z) & \bE\{\rho(Z)\}\le D\end{array}\right.\nonumber\\
&=&\sup_{\{Z:~\bE\{\rho(Z)\}\le D\}} h(Z),
\end{eqnarray}
where (a) is due to the fact that the objective function, $-\int_{\calU}\mbox{d}z
f(z)\log f(z)
+\beta\left[D-\int_{\calU} f(z)\rho(z)\mbox{d}z\right]$, is concave in $f$ and
affine (and hence convex) in $\beta$.

\end{document}